\documentclass[pre]{revtex4}
\usepackage{psfrag}
\usepackage{amssymb,amsmath,amsthm}
\usepackage[dvips]{graphicx}
\usepackage{verbatim}

\begin{document}
\title{VARIATIONAL FORMULATION FOR THE KPZ AND RELATED KINETIC
EQUATIONS}

\author{Horacio S. Wio }

\affiliation{Instituto de F\'{\i}sica de Cantabria, Universidad de
Cantabria \& CSIC, E-39005 Santander, Spain}

\begin{abstract}

\noindent We present a variational formulation for the
Kardar-Parisi-Zhang (KPZ) equation that leads to a
thermodynamic-like potential for the KPZ as well as for other
related kinetic equations. For the KPZ case, with the knowledge of
such a potential we prove some global shift invariance properties
previously conjectured by other authors. We also show a few results
about the form of the stationary probability distribution function
for arbitrary dimensions. The procedure used for KPZ was extended in
order to derive more general forms of such a functional leading to
other nonlinear kinetic equations, as well as cases with density
dependent surface tension.

\bigskip


\end{abstract}

\maketitle

\noindent {\bf 1. Introduction} \smallskip

\noindent Phenomena far from equilibrium are ubiquitous in nature,
including among many other, turbulence in fluids, interface and
growth problems, chemical reactions, biological systems, as well as
economical and sociological structures. During the last decades the
focus on statistical physics research has shifted towards the study
of such systems. Among those studies, the understanding of growing
kinetics at a microscopic as well as on a mesoscopic level
constitutes a major challenge in physics and material science [Tong
\& Williams, 1994; Barab\'asi \& Stanley, 1995; Halpin-Healy \&
Zhang, 1995; Marsili {\it et al.}, 1996]. Some recent papers have
shown how the methods and know-how from static critical phenomena
have been exploited within nonequilibrium phenomena of growing
interfaces, obtaining scaling properties, symmetries, morphology of
pattern formation in a driven state, etc [Hentschel, 1994;
Pr\"{a}hofer \& Spohn, 2004; L\'opez {\it et al.}, 2005; Fogedby,
2006; Ma {\it et al.}, 2007; Castro {\it et al.}, 2007].

\noindent Even though it was (briefly) discussed in [Cross \&
Hohenberg, 1993], there is still a common belief that the nontrivial
spatial-temporal behavior occurring in several nonequilibrium
systems, originates from the {\it non-potential} (or
\textit{non-variational}) character of the dynamics, meaning that
there is no Lyapunov functional for the dynamics. However, Graham
and co-workers have shown in a series of papers [Graham, 1987;
Graham \& Tel, 1990] that a Lyapunov-like functional exists for a
very general dynamical system, the complex Ginzburg-Landau equation.
Such a functional is formally defined as the solution of a
Hamilton-Jacobi-like equation, or obtained in a small gradient
expansion [Descalzi \& Graham, 1994]. The confusion associated with
the qualification of \textit{nonvariational} dynamics comes from the
idea that the dynamics of systems having nontrivial attractors
(limit cycle, chaotic) cannot be deduced from the minimization of a
potential playing the same role as the free energy in equilibrium
systems [Cross \& Hohenberg, 1993]. Nevertheless, this does not
preclude the existence of a Lyapunov functional for the dynamics
that will have local minima identifying the attractors of the system
[Montagne {\it et al.}, 1996; Wio, 1997; Wio {\it et al.}, 2002; Wio
\& Deza, 2007]. However, once the system has reached an attractor
that is not a fixed point, the dynamics proceeds inside the
attractor driven by \textit{nonvariational} contributions to the
dynamical flow, that do not change the value of the Lyapunov
functional, implying that the dynamics is not completely determined
once the indicated functional is known. This situation has known
examples even in equilibrium statistical mechanics [Hohenberg \&
Halperin, 1977]. Hence, the Lyapunov functional, or
\textit{nonequilibrium potential} (NEP) [Graham, 1987; Graham \&
Tel, 1990; Wio, 1997; Ao, 2004], plays the role in nonequilibrium
situations of a thermodynamical-like potential characterizing the
global properties of the dynamics: attractors, relative (or
nonlinear) stability of these attractors, height of the barriers
separating attractions basins, offering the possibility of studying
transitions among the attractors due to the effect of (thermal)
fluctuations.

\noindent In a recent series of papers we have shown several results
related to the obtention of the indicated NEP's for other system's
classes:  scalar and non-scalar reaction-diffusion systems [Bouzat
\& Wio, 1999; von Haeften {\it et al.}, 2004; von Haeften {\it et
al.}, 2005; von Haeften \& Wio, 2007]. In particular we have
exploited those results for the study of stochastic resonance
[Gammaitoni {\it et al.}, 1998] in extended systems [Wio, 1997;
Bouzat \& Wio, 1999; von Haeften {\it et al.}, 2000; von Haeften
{\it et al.}, 2004; von Haeften {\it et al.}, 2005; von Haeften \&
Wio, 2007; Tessone \& Wio, 2007; Wio \& Deza, 2007;]. In those
works, we have analyzed problems of stochastic resonance in scalar
and activator-inhibitor systems, in systems with local and nonlocal
interactions, studied system-size stochastic resonance, etc.

\noindent Here, and related to the kinetics of growing interfaces,
we discuss the case of the Kardar-Parisi-Zhang equation (KPZ)
[Kardar, Parisi \& Zhang, 1986; Medina {\it et al.}, 1989]. This
equation, that describes the evolution of $h(\bar{x},t)$, a field
that corresponds to the height of a fluctuating interface, reads
\begin{eqnarray}\label{eq-000}
\frac{\partial h(\bar{x},t)}{\partial t} = \nu \nabla^2 h(\bar{x},t)
+ \frac{\lambda}{2} \left(\nabla h(\bar{x},t) \right)^2 + K_o + \xi
(\bar{x},t),
\end{eqnarray}
where $\xi (\bar{x},t)$ is a Gaussian white noise, of zero mean
($\langle \xi (\bar{x},t) \rangle = 0$) and correlation $\langle \xi
(\bar{x},t) \xi (\bar{x}',t') \rangle = 2 \varepsilon \delta
(\bar{x}-\bar{x}') \delta (t-t')$. As indicated above, this
nonlinear differential equation describes fluctuations of a growing
interface with a surface tension given by $\nu$, $\lambda$ is
proportional to the average growth velocity and arises because the
surface slope is parallel transported in such a growth process.

\noindent Opposing to a claim in a recent paper [Fogedby,
2006]:$\,\,$ \textit{The KPZ equation is in fact a genuine kinetic
equation describing a nonequilibrium process in the sense that the
drift $\nu \nabla^2 h + \frac{\lambda}{2} \nabla h \cdot\nabla h -
F$ cannot be derived from an effective free energy};$\,\,$ in [Wio,
2007] it was shown that such a \textbf{\textit{nonequilibrium
thermodynamic-like potential}} (NETLP) for the KPZ equation exists.

\noindent In this paper we present the approach to obtain the NETLP
for the KPZ equation, and also show how it is possible to obtain
such a NETLP for other related kinetic equations, as well as extend
the procedure to other general situations. The organization is as
follows. In the next Section we show how to obtain the NETLP for the
KPZ case and, exploiting its knowledge, we discuss conjectures
advanced in [Hentschel, 1994] and how they are fulfilled. Next we
discuss about the form of the stationary (or asymptotic) probability
distribution function, \textit{\textbf{valid for any dimension}}. In
the following Section we discuss the situation with a non-local
interaction and symmetric kernels, showing how this form of
extending the derivation procedure allows to obtain more general
forms of kinetic equations. The following Section discusses the case
of density dependent surface tension, while the last one presents
some final comments.

\bigskip

\noindent {\bf 2. Kardar-Parisi-Zhang case} \smallskip

\smallskip

\noindent {\bf 2.1 Derivation}  \smallskip

\noindent In order to show how to obtain the above indicated
functional, we start considering the following simple (general)
scalar reaction-diffusion equation with multiplicative noise
\begin{equation}\label{eq-01}
\frac{\partial}{\partial t} \phi (\bar{x},t) = \nu \nabla^2 \phi
(\bar{x},t) + f(\phi (\bar{x},t)) + \phi (\bar{x},t) \eta
(\bar{x},t),
\end{equation}
where $f(\phi (\bar{x},t))$ is a general nonlinear function. The
noise $\eta (\bar{x},t)$ is a Gaussian white noise, of zero mean and
intensity $\sigma$. For this equation we assume the Stratonovich
interpretation.

\noindent It is known that the system in Eq. (\ref{eq-01}) has the
following NETLP
\begin{eqnarray}\label{eq-02}
{\cal F}[\phi] = \int_{\Omega} \left\{ - \int ^{\phi (\bar{x},t)}
_{0} f (\varphi) \, d\varphi + \frac{\nu}{2} \left( \nabla \phi
(\bar{x},t) \right)^2 \, \right\} d\bar{x},
\end{eqnarray}
where $\Omega$ indicates the integration range, and
\begin{eqnarray}\label{eq-03}
\frac{\partial}{\partial t} \phi (\bar{x},t) = - \frac{\delta {\cal
F}[\phi]}{\delta \phi (\bar{x},t)} + \phi (\bar{x},t) \, \eta
(\bar{x},t);
\end{eqnarray}
where the contribution from the boundaries is null, due to the
variation $\delta\phi$ being fixed ($=0$) at these boundaries, as
usual. As has been shown in previous works [Iz\'us {\it et al.},
1998; Bouzat \& Wio, 1998], it also fulfills the Lyapunov
characteristic $\frac{\partial}{\partial t}{\cal F}[\phi] \leq 0$
(it is worth here commenting that, as a matter of fact, this
conditions is only valid in a weak noise limit).

\noindent Exploiting the so called Hopf-Cole transformation we now
define a new field, $h(\bar{x},t)$ that, as indicated before,
corresponds to an interface height,
\begin{equation}\label{eq-04}
h(\bar{x},t) = \frac{2 \nu}{\lambda}\ln \phi (\bar{x},t),
\end{equation}
with the inverse $$\phi (\bar{x},t) = e^{\frac{\lambda}{2 \nu}
h(\bar{x},t)}.$$ As $\phi (\bar{x},t) \geq 0$, $h(\bar{x},t)$ is
always well defined. The transformed equation reads
\begin{equation}
\label{eq-04p}
\partial_t h(\bar{x},t) = \nu \nabla^2 h
+ \frac{\lambda}{2} \left(\nabla h \right)^2 + \frac{\lambda}{2 \nu}
e^{- \frac{\lambda h}{2 \nu}} \hat{f}(h) + \xi (\bar{x},t),
\end{equation}
where $\hat{f}(h)=f(\phi)=f(e^{\frac{\lambda}{2 \nu} h})$. Now, in
order to reduce our general result to the one shown in Eq.
(\ref{eq-04p}), we assume $f(\phi (\bar{x},t)) = a \phi (\bar{x},t)$
(with $a$ some constant). In this way Eq. (\ref{eq-01}) becomes Eq.
(\ref{eq-000}), with $a = \frac{\lambda}{2 \nu} K_o$ and $\sigma =
\left( \frac{\lambda}{2 \nu}\right) ^2 \varepsilon$. However, the
noise term that in the original equation (Eq. (\ref{eq-01})) has a
multiplicative character, in the transformed equation (Eq.
(\ref{eq-000})) becomes additive.

\noindent If we now apply the same transformation to the NETLP
indicated in Eq.(\ref{eq-02}), restricting ourselves to $f(\phi
(\bar{x},t)) = a \phi (\bar{x},t)$, we obtain
\begin{eqnarray}\label{eq-001}
{\cal F}[h] = \int_{\Omega} e^{\frac{\lambda}{\nu} h(\bar{x},t)}
\frac{\lambda}{2 \nu} \Bigl[ - K_o + \frac{\lambda }{4} \left(
\nabla h(\bar{x},t) \right)^2 \Bigr] d\bar{x}.
\end{eqnarray}
It is easy to prove that this functional fulfills both, the relation
\begin{eqnarray}\label{eq-002}
\frac{\partial}{\partial t} h(\bar{x},t) & = & - \Gamma [h]
\frac{\delta {\cal F}[h]}{\delta h(\bar{x},t)} + \xi (\bar{x},t);
\end{eqnarray}
as well as the Lyapunov characteristic $\frac{\partial}{\partial
t}{\cal F}[h] \leq 0$, where the function $\Gamma [h]$ is given by
$$ \Gamma [h] = \left(\frac{2 \nu}{\lambda} \right)^2 e^{-
\frac{\lambda}{\nu} h(\bar{x},t)}.$$ Hence we have a \textbf{free
energy-like functional} from where the KPZ kinetic equation can be
obtained through functional derivation. Clearly, the contribution to
the variation coming from the boundaries is again null.

\noindent It is worth to consider once more Eq. (\ref{eq-01}), but
now assuming $f(\phi (\bar{x},t)) = a \phi (\bar{x},t) - b \, \phi
(\bar{x},t)^3$, i.e. we include a limiting (or saturating) term. In
such a case, the reaction-diffusion equation corresponds to a form
of the so called Schl\"{o}gl model [Mikhailov, 1990; Wio, 1994]. The
``potential" contribution to the NETLP in Eq. (\ref{eq-02}) has the
form
$$- \int_{\Omega} \left( \frac{a}{2} \phi (\bar{x},t)^2 -
\frac{b}{4} \phi (\bar{x},t)^4 \right)\, d\bar{x}.$$ Applying again
the indicated Hopf-Cole transformation, in Eq. (\ref{eq-000}) a new
associated term arises, having the form $ - \gamma \,
e^{\frac{\lambda}{\nu} h(\bar{x},t)},$ with $b = \frac{\lambda}{2
\nu} \gamma$. The new kinetic equation corresponds to a form of the
so called \textit{bounded-KPZ} [Grinstein {\it et al.}, 1996; Tu
{\it et al.}, 1997; de los Santos {\it et al.}, 2007]. Clearly, we
will also have an extra term in the associated NETLP (Eq.
(\ref{eq-001})). However, here we restrict ourselves to the case $b
= 0$, and only analyze the more ``usual" form of the KPZ equation
indicated by Eq. (\ref{eq-000}).

\bigskip

\noindent {\bf 2.2 Some Properties}  \smallskip

\noindent Let us now check some of the properties assumed previously
for such a functional. According to the analysis of global shift
invariance in [Hentschel, 1994], it is easy to see that the
relations indicated by Eq. (9) in the indicated paper are fulfilled.
That is, we can prove that if $l$ is an arbitrary (constant) shift
\begin{eqnarray}\label{eq-08}
{\cal F}[h + l]& = & K[l] {\cal F}[h]\nonumber \\
\Gamma [h + l] & = & K[l]^{-1} \Gamma [h],
\end{eqnarray}
with $ K[l] = e^{\frac{\lambda}{D} l} = \left( \frac{2 \nu}{\lambda}
\right)^2 \Gamma [l]^{-1}$.

\noindent To prove other conjectures also indicated in [Hentschel,
1994], we introduce the \textit{free energy-like density}
$\widetilde{{\cal F}}[h,\nabla h],$ defined through
$${\cal F}[h] = \int d \bar{x} \, \widetilde{{\cal F}}[h,\nabla
h].$$ Hence, $\widetilde{{\cal F}}[h,\nabla h] = \frac{\lambda}{2
\nu} e^{\frac{\lambda}{\nu} h(\bar{x},t)}  [ - K_o + \frac{\lambda
}{4} \left( \nabla h(\bar{x},t) \right)^2 ].$ The relations we refer
are
\begin{eqnarray}\label{eq-08p}
\widetilde{{\cal F}}[h, \nabla h]& = & e^{sh} \widetilde{{\cal F}}_1
[(\nabla h)^2] \nonumber \\
\Gamma [h, \nabla h] & = & e^{-sh} \Gamma _1 [(\nabla h)^2],
\end{eqnarray}
and according to the form of $\widetilde{{\cal F}}[h,\nabla h]$, it
is clear that the first relation above results obviously true, while
for the second relation we have that $\Gamma [h, \nabla h] = e^{- s
h(\bar{x},t)} \Gamma_o,$ where $\Gamma_o=1$, and
$s=\frac{\lambda}{\nu}$, as $\Gamma [h]$ is not function of $\nabla
h$. In addition, it can be also proved that the indicated NETLP is
also invariant under the nonlinear Galilei transformation that, as
discussed in [Fogedby, 2006], are fulfilled by the KPZ equation.

\noindent It is here adequate to make a warning. We have found that,
from the indicated \textit{free energy}-like functional for the KPZ
kinetic equation and by a functional derivative, we can obtain a
form that resembles a (relaxation) model \textbf{A} according to the
classification in [Hohenberg \& Halperin, 1977]. In one hand, in the
standard ``model A" it is known that the dynamics can be seen as a
superposition of modes that decay exponentially towards a steady
state. In addition, it is also known that the time-dependent
correlations obey certain constraints such as positivity. On the
other hand, in the KPZ problem it is known that the relaxation of
perturbations decay in a stretched exponential manner [Schwartz \&
Edwards, 2002; Edwards \& Schwartz, 2002; Colaiori \& Moore, 2002;
Katzav \& Schwartz, 2004]. Hence, even though Eq. (\ref{eq-002})
looks similar to ``model A", its behavior is far from trivial, and
we have no a-priori intuition to what its dynamic could be. Clearly,
this is a point to have in mind when suggesting any kind of Antsatz
for the temporal behavior.

\bigskip

\noindent {\bf 3. Probability Distribution Function for KPZ}
\smallskip

\noindent {\bf 3.1 General Aspects}  \smallskip

\noindent The knowledge of the NETLP indicated in the previous
section, allows us to readily write the asymptotic long time
probability distribution (pdf). We start writing the (functional)
Fokker-Planck equation (FPE) associated to Eq. (\ref{eq-000}), that
reads
\begin{eqnarray}\label{pdf-01}
\frac{\partial}{\partial t}\mathcal{P}[h(\bar{x},t)] = \int
_{\Omega} d \bar{x} \frac{\delta}{\delta h} \Bigl\{ - \Bigl[ \nu
\nabla^2 h(\bar{x},t) + \frac{\lambda}{2} \left(\nabla
h(\bar{x},t)\right)^2 \Bigr] \mathcal{P}[h(\bar{x},t)] +
\varepsilon\, \frac{\delta }{\delta h} \mathcal{P}[h(\bar{x},t)]
\Bigr\};
\end{eqnarray}
where, in order to focus in the most relevant aspects and to
simplify, we adopted $K_o = 0.$

\noindent In Chap. 6 of [Barab\'asi \& Stanley, 1995] as well as in
Sect. 3.5 of [Halpin-Healy \& Zhang, 1995] the form of the
probability distribution function (pdf) for the one dimensional KPZ
equation was discussed. The form of such a pdf for an arbitrary
dimension, so far, is not known. However, here we show the general
form of such a pdf.

\noindent The knowledge of such a NETLP for the KPZ equation allows
us to readily write the asymptotic long time probability
distribution function (pdf), valid for (worth to be remarked)
\textbf{any dimension}, which (due to the ``diagonal" character of
$\Gamma [h]$) is given by
\begin{eqnarray}\label{eq-30}
{\cal P}_{as}[h(\bar{x})] & \sim & \exp \left\{ - \frac{1}
{\varepsilon} \int d\bar{x} \, \int ^{h(\bar{x})} _{h_{ref}} d\psi
\, \Gamma [\psi] \frac{\delta {\cal F}[\psi]}{\delta \psi} \right\}
\nonumber \\
& \sim & \exp \Bigl\{ - \frac{1}{\varepsilon} \int d\bar{x} \,
\Bigl( \Gamma [h] \widetilde{{\cal F}}[h] - \int ^{h(\bar{x})}
_{h_{ref}} d \psi \, \frac{\delta \Gamma
[\psi]}{\delta \psi} \widetilde{{\cal F}} [\psi] \Bigr) \Bigr\} \nonumber \\
& \sim & \exp \Bigl\{ - \frac{\nu}{2 \varepsilon} \int d\bar{x}
\left(\nabla h \right)^2 + \frac{\lambda}{2 \varepsilon} \int
d\bar{x} \int ^{h(\bar{x})}_{h_{ref}} d\psi \left(\nabla \psi
\right)^2 \Bigr\} \nonumber \\
& \sim & \exp \left\{- \frac{\Phi [h]}{\varepsilon} \right\},
\end{eqnarray}
where, we reiterate, we assumed $K_o = 0$, and ${h_{ref}}$ is an
arbitrary reference state. The second line results by using
functional methods (see for instance [H\"{a}nggi, 1985]). The third
line shows a nice structure, where we can identify a contribution,
with a Gaussian dependence on the slope, plus a ``correction" term
proportional to $\lambda .$ The above indicated result is the valid
(albeit ``formal") solution for arbitrary dimension irrespective of
boundary conditions, while for the one-dimensional case, and the
adequate (periodic) boundary conditions, it is possible to show that
this pdf reduces to the well known Gaussian result [Barab\'asi \&
Stanley, 1995; Halpin-Healy \& Zhang, 1995].

\noindent Clearly, a relevant point is related to the interpretation
of the integral over the function $\psi (\bar{x})$ in Eq.
(\ref{eq-30}). In order to simplify we consider the 1-d case, and
focus only on the first term. Using functional techniques, we have
\begin{eqnarray} \label{app-01}
\nu \int d{x} \int ^{h({x})} _{h_{ref}} d \psi \,\frac{\partial ^2
}{\partial x^2} \psi (x) = - \frac{\nu}{2} \int d{x} \int ^{h({x})}
_{h_{ref}} d \psi \frac{\delta}{\delta \, \psi (x)} \left( \int
d{x'} \left(\frac{\partial \psi (x') }{\partial x'} \right)^2
\right) =  \frac{\nu}{2} \int d{x} \left( \frac{\partial h
}{\partial x} \right)^2,
\end{eqnarray}
where we have used that $\frac{\delta \,\psi (x')}{\delta \, \psi
(x)} = \delta (x-x')$. Another way to grasp it is via a discrete
representation
\begin{eqnarray} \label{app-02}
& \approx & \sum_{j}\int ^{h_{j}} d \psi_{j} \, \left( \psi_{j+1} -
2 \psi_{j} + \psi_{j-1} \right) \nonumber\\ & \approx &
\frac{\nu}{2} \sum_{j} \int ^{h_{j}} d \psi_{j} \, \frac{\partial
}{\partial \psi_{j}} \sum_{l} \, [\psi_{l+1} -  \psi_{l}]^2,
\nonumber
\end{eqnarray}
that, as only the $l=j$ and $l = j-1$ terms survive, yields
\begin{eqnarray} \label{app-03}
\approx \frac{\nu}{2}\, \sum_{j} \, [h_{j+1} -  h_{j}]^2.
\end{eqnarray}
In both cases we obtain the known result.

\noindent The interpretation for the second (KPZ) term is analogous
to the one in the example indicated above. However, for this case the
situation is a little more delicate. Several recent papers have
discussed different alternatives to the discrete form of such a
contribution in the KPZ equation in 1-d [Newman \& Bray, 1996;
Lam \& Shin, 1998; Ma, Jiang \& Yang, 2007]. We will not come into
these details here, that will be deeply discussed in [Revelli \& Wio,
2008].

\bigskip

\noindent {\bf 3.2 Nonequilibrium Potential for KPZ}
\smallskip

\noindent The last line in Eq. (\ref{eq-30}) defines the functional
$\Phi [h]$
\begin{eqnarray} \label{nep-01}
\Phi [h] = \frac{\nu}{2} \int d\bar{x} \left(\nabla h \right)^2 - \,
\frac{\lambda}{2} \int d\bar{x} \int ^{h(\bar{x})} _{h_{ref}} d
\psi \, \left(\nabla \psi \right)^2.
\end{eqnarray}
The variation of this functional give us
\begin{eqnarray}
\frac{\partial}{\partial \, t} h(\bar{x},t) = - \frac{\delta \Phi
[h]}{\delta h(\bar{x},t)} + \xi(\bar{x},t).
\end{eqnarray}
This functional also fulfills the Lyapunov condition
$\frac{\partial}{\partial \, t} \Phi [h] = - \left(\frac{\delta \Phi
[h]}{\delta h(\bar{x},t)} \right)^2 \leq 0$.

\noindent Hence, such a functional could be identified as the
\textit{nonequilibrium potential} [Graham, 1987; Graham \& Tel,
1990; Wio {\it et al.}, 2002] for the KPZ case. More, even though it
could be also identified with a Hamiltonian for KPZ, it is worth to
remark that it differs from the form indicated in Eq. (3.4) of
[Halpin-Healy \& Zhang, 1995].

\noindent The last functional form (albeit so far only formal)
could, in principle, allow us to exploit several known techniques
[Graham, 1987; Graham \& Tel, 1990; Wio {\it et al.}, 2002]. It also
shows that the claim by previous authors indicated at the
Introduction is not true.

\bigskip

\noindent {\bf 3.3 NEP's application: A simple example}
\smallskip

\noindent Here we discuss a simple example in order to show the
possibilities that offers the knowledge of such a NETLP. For this
example we analyze a sightly different situation than the one
studied in Eq. (\ref{eq-000}) and its associated NEP in Eq.
(\ref{nep-01}). We only consider a spatial quenched noise
(``disorder") instead of the spatial-temporal noise considered so
far. The equation associated to such a problem is
\begin{eqnarray}\label{eq-100}
\frac{\partial}{\partial t} h(\bar{x},t) = \nu \nabla^2 h(\bar{x},t)
+ \frac{\lambda}{2} \left( \nabla h(\bar{x},t) \right)^2 +  K_o +
\vartheta (\bar{x}),
\end{eqnarray}
where, as in previous studies [Nattermann \& Renz, 1989; Krug \&
Halpin-Healy, 1993; Ramasco, L\'opez \& Rodriguez, 2006; Szendro,
L\'opez \& Rodriguez, 2007], $\vartheta (\bar{x})$ is a quenched,
Gaussian distributed, noise. For this case we have that
\begin{eqnarray}\label{eq-101}
\frac{\partial}{\partial t} h(\bar{x},t) = - \Gamma [h] \frac{\delta
{\cal F}[h]}{\delta h(\bar{x},t)},
\end{eqnarray}
with the same form of ${\cal F}[h]$ as in Eq. (\ref{eq-001}), but
$K_o$ replaced by $K_o + \vartheta (\bar{x})$.

\noindent The indicated studies have shown that the front profile
presents a triangular structure [Ramasco, L\'opez \& Rodriguez,
2006; Szendro, L\'opez \& Rodriguez, 2007] and, clearly, it is of
relevance to determine the slope of such structures. Using the known
form of the NETLP, we can minimize this free-energy-like  functional
and obtain, in the 1-d case, that such a slope is $\alpha = \left(
\frac{16}{\nu \lambda}\right)^{1/3},$ a value that agrees quite well
with the numerical evaluations.

\bigskip

\noindent {\bf 4. Other Kinetic Equations: Non Locality}
\bigskip

\noindent We can go still further and look for the possibility of
deriving a NETLP for more general forms of kinetic equations. To
this end, let us assume that we have the following non-local
reaction-diffusion equation [Wio {\it et al.}, 2002]
\begin{eqnarray}\label{eq-21}
\frac{\partial}{\partial t} \phi (\bar{x},t) = & \nu & \nabla^2 \phi
(\bar{x},t)+ \, a \phi (\bar{x},t) - \beta \int_{\Omega} d\bar{x'}
\mathbf{G}(\bar{x},\bar{x'}) \phi (\bar{x'},t) + \phi
(\bar{x},t)\,\eta (\bar{x'},t),
\end{eqnarray}
where, as discussed in [Bouzat \& Wio, 1999; von Haeften {\it et
al.}, 2004; von Haeften \& Wio, 2007], the kernel
$\mathbf{G}(\bar{x}, \bar{x'})$ could be of a very general
character, and $\beta$ is the interaction intensity. It was shown
that the form of the associated NETLP is
\begin{eqnarray}\label{eq-22}
{\cal F}[\phi] = \int_{\Omega} \Bigl[ - \frac{a}{2} \phi
(\bar{x},t)^2  + \frac{\nu}{2} \left( \nabla \phi (\bar{x},t)
\right)^2 + \beta \int_{\Omega} d\bar{x'} \, \phi (\bar{x},t)
\,\mathbf{G}(\bar{x},\bar{x'})\, \phi (\bar{x'},t) \Bigr] d\bar{x}.
\end{eqnarray}

\noindent As we have done before, using the Hopf-Cole transformation
we obtain a \textit{generalized-non-local} form of the KPZ equation
\begin{eqnarray}\label{eq-25}
\frac{\partial}{\partial t} h(\bar{x},t) = \nu \nabla^2 h(\bar{x},t)
+ \frac{\lambda}{2} \left( \nabla h(\bar{x},t) \right)^2 + K_o -
\beta e^{-\frac{\lambda}{2 \nu} h(\bar{x},t)} \int_{\Omega}
d\bar{x'} \mathbf{G}(\bar{x},\bar{x'}) e^{\frac{\lambda}{2 \nu}
h(\bar{x'},t)} + \xi (\bar{x},t).
\end{eqnarray}
Even though the nonlocal contribution indicated above differs from
those discussed in [Mukherji \& Bhattacharjee, 1997; Katzav, 2003],
it is clear that such form of a nonlocal term is also of great
interest. Repeating the previous procedure we find the associated
NETLP
\begin{eqnarray}\label{eq-26}
{\cal F}[h] = \int_{\Omega} d\bar{x} e^{\frac{\lambda}{\nu}
h(\bar{x},t)} \Bigl[ - \frac{\lambda}{2 \nu} K_o +
\left(\frac{\lambda ^2}{8 \nu}\right) \left( \nabla h(\bar{x},t)
\right)^2 + \beta e^{- \frac{\lambda}{2 \nu} h(\bar{x},t)}
\int_{\Omega} d\bar{x'} \mathbf{G}(\bar{x}, \bar{x'})
e^{\frac{\lambda}{2 \nu} h(\bar{x'},t)} \Bigr].
\end{eqnarray}
At this stage it is required to make some assumptions about the
kernel.

\noindent We assume that the nonlocal kernel has translational
invariance, that is $\mathbf{G}(\bar{x}, \bar{x'}) =
\mathbf{G}(\bar{x} - \bar{x'}).$ Also, that it is of (very) ``short"
range, that allows us to expand it as
\begin{equation} \label{expand} \mathbf{G}(\bar{x} - \bar{x'}) =
\sum_{n=0}^{\infty} A_{2n} \delta^{(2n)} (\bar{x}-\bar{x'}),
\end{equation}
with $\delta^{(n)} (\bar{x}-\bar{x'}) = \nabla _{x'}^{n}\delta
(\bar{x}-\bar{x'})$, and symmetry properties taken into account.
Exploiting this form of the kernel, we arrive to the following
contributions in Eq. (\ref{eq-25})
\begin{eqnarray}\label{eq-27}
e^{-\frac{\lambda}{2 \nu} h(\bar{x},t)} \beta \int_{\Omega}
d\bar{x'} \mathbf{G}(\bar{x} - \bar{x'}) e^{\frac{\lambda}{2 \nu}
h(\bar{x'},t)} &\approx& \beta \left\{ A_0 + A_2 \left[ \left(
\frac{\lambda}{2 \nu} \right)^2 \left( \nabla h
\right)^2 + \frac{\lambda}{2 \nu} \nabla^2 h \right] \right. \nonumber \\
& & + A_4 \left[ \left( \frac{\lambda}{2 \nu} \right)^4 \left(\nabla
h \right)^4 + 6 \left( \frac{\lambda}{2 \nu} \right)^3
\left( \nabla h \right)^2 \nabla^2 h \right. \nonumber \\
&& \,\, \left. \left. + 2\, \left( \frac{\lambda}{2 \nu} \right)^2
\nabla ^2 \left( \nabla h \right)^2 - \left( \frac{\lambda}{2 \nu}
\right)^2 \left( \nabla^2 h \right)^2 + \, \frac{\lambda}{2 \nu}\,
\, \nabla^4 h \right] \right\} + A_6 \,\ldots .,
\end{eqnarray}
where the last term indicates contributions of order $n \geq 3$
($2n=6$). The parameter $\beta$ could (in principle) be positive or
negative, indicating an inhibitor or an activator role for the
nonlocal interaction term, respectively.

\noindent These contributions have the same form of those ones that
arose in several previous works, where scaling properties, symmetry
arguments, etc, have been used to discuss the possible contributions
to a general form of the kinetic equation [Hentschel, 1994; Linz
{\it et al.}, 2000; Lopez {\it et al.}, 2005]. Clearly, the different
contributions that arose in Eq. (\ref{eq-27}) are tightly related to
several of other previously studied equations, like the
Kuramoto-Sivashinsky [Sivashinsky, 1977; Kuramoto, 1978], the
Sun-Guo-Grant equation  [Sun {\it et al.}, 1989], and others
[Hentschel, 1994].

\bigskip

\noindent {\bf 5. Density Dependent Surface Tension}

\bigskip

\noindent We can also go further in another direction. Let us
consider the case of \textit{density dependent diffusion (surface
tension)}. Following the same original approach, we start
considering the following reaction-diffusion equation with a density
dependent diffusion for a field $\phi (x,t)$, as studied in [von
Haeften {\it et al.}, 2000; Tessone \& Wio, 2007]
\begin{equation}
\label{Ballast}
\partial_t\phi(x,t)=\nabla \left( \nu(\phi) \nabla \phi
\right) + f(\phi) + \phi(x,t) \eta (x,t).
\end{equation}
As before, $f(\phi)$ is a general nonlinear function, that we
collapse to $f(\phi) = a \phi(x,t).$ As in Eq. (\ref{eq-01}), $\eta
(x,t)$ is a delta correlated white noise of intensity $\sigma$. We
assume $\nu(\phi) = \nu_o g(\phi),$ that --using the Hopf-Cole
transformation-- yields $\tilde{\nu}(h) = \nu_o \tilde{g}(h).$

\noindent The transformed equation reads
\begin{equation}
\label{Ballast2}
\partial_t h(x,t) = \nabla \left( \tilde{\nu}(h) \nabla h \right)
+ \frac{\tilde{\lambda} (h)}{2} \left(\nabla h \right)^2 + K_o + \xi
(x,t),
\end{equation}
with $\tilde{\lambda} (h) = \lambda _o \tilde{g}(h)$ and $a =
\frac{\lambda _o}{2 \nu _o} K_o.$ This equation has a KPZ-like form,
with both a density dependent surface tension term and a density
dependent nonlinear coupling, with both nonlinearities having the
same functional dependence.

\noindent In order to see the associated NETLP, let us remember its
form for the original reaction-diffusion equation (Eq.
(\ref{Ballast})), it reads
\begin{equation} \label{Ballast3}
{\cal F}[\phi]=\int_{\Omega} \left\{ -\int_0 ^{\phi} \nu(\phi')
f(\phi')\,d\phi' + \frac{1}{2} \left( \nu(\phi) \nabla
\phi\right)^2\,\right\} \, dx,
\end{equation}
that fulfills [von Haeften  {\it et al.}, 2000]
$$\frac{\partial}{\partial t}\phi(x,t) = -\frac{1}{\nu(\phi)}
\frac{\delta {\cal F}[\phi]}{\delta \phi} + \phi \, \eta (x,t).$$

\noindent Applying once more the Hopf-Cole transformation, we obtain
\begin{eqnarray} \label{Ballast4}
{\cal F}[h] = \int_{\Omega}  dx \, \left( \frac{\lambda _o}{2 \nu
_o} \right) ^2 \Bigl\{ - \int ^{h} du \, e^{\frac{\lambda _o u}{2
\nu_o}} \tilde{\nu}(u)\, K_o + \frac{1}{2} \,\left( \tilde{\nu}(h)
e^{\frac{\lambda _o h}{2 \nu_o}} \nabla h \right)^2 \Bigr\}.
\end{eqnarray}

\noindent From this NETLP we can prove that
\begin{eqnarray}\label{Ballast5} \frac{\partial}{\partial t}
h(\bar{x},t) & = & - \tilde{\Gamma}[h] \frac{\delta {\cal
F}[h]}{\delta h(x,t)} + \xi (x,t);
\end{eqnarray}
where $\xi (x,t)$ is a white noise of intensity $\varepsilon$
(with $\sigma = \left( \frac{\lambda _o}{2 \nu _o} \right) ^2
\varepsilon$), and the function $\tilde{\Gamma} [h]$ is given by
$$\tilde{\Gamma} [h] = \left(\frac{2 \nu_o}{\lambda _o} \right)^2
\frac{e^{- \frac{\lambda_o}{\nu_o} h(\bar{x},t)}}{\tilde{\nu}(h)}.$$
As before, if we have $f(\phi (\bar{x},t)) = a \phi (\bar{x},t) - b
\, \phi (\bar{x},t)^3$, we get a generalized form of the ``bounded
KPZ" equation.\\

\bigskip

\noindent {\bf 6. Conclusions}  \smallskip

\noindent The previous results indicates that, for a very general
form of a KPZ-like equation, we can device a NETLP \textit{\`{a} la
carte}. Consider the following equation
\begin{equation} \label{KPZ-01}
\partial_t h(\bar{x},t) = \nu \nabla^2 h +
\frac{\lambda}{2} \left(\nabla h \right)^2 + F(h) + \xi (\bar{x},t)
\end{equation}
where $F(h)$ is a general nonlinear function of $h$. According to
the previous results we could readily write the associate NETLP that
has the form
\begin{eqnarray}
\label{KPZ-02} {\cal F}[h]= \int \left\{ - \frac{\lambda}{2 \nu}
\int ^{h} _{h_{ref}} e^{\frac{\lambda u}{\nu}} F(u) du  +
\frac{\lambda ^2 }{8 \nu} \left( e^{\frac{\lambda h}{2 \nu}} \nabla
h \right)^2 \right\} dx.
\end{eqnarray}
Clearly, it fulfills the Lyapunov condition
$\frac{\partial}{\partial t} {\cal F}[h(\bar{x},t)] \leq 0,$ and
also, through functional derivation, allows us to obtain the kinetic
equation as
\begin{equation} \label{KPZ-03}
\frac{\partial}{\partial t} h(\bar{x},t) = - \Gamma [h] \frac{\delta
{\cal F}[h]}{\delta h(x,t)} + \xi (x,t),
\end{equation}
with $\Gamma [h] = \left(\frac{2 \nu}{\lambda } \right)^2 e^{-
\frac{\lambda}{\nu} h(\bar{x},t)}.$ Hence, it is clear that we can
write the NETLP for a very general KPZ-like form, independently of
the actual form of $F(h)$.

\noindent Summarizing, we have here found the form of the Lyapunov
functional or NETLP for the KPZ equation. More, we have devised a
way to extent the procedure to derive it, and in such a way we were
able to derive more general forms, including several kinetic
equations studied in the literature of interface growing phenomena.
Even the case of density dependent surface tension, have been
discussed. From this NETLP, and through a functional derivative, we
have obtained either the KPZ as well as other generalized kinetic
equations. We have also shown that the NETLP for KPZ fulfills global
shift properties, as well as other ones anticipated for such an
unknown functional. More, we have found the exact expression, valid
for any dimension, for the probability distribution function, and
have commented on the result of a simple example that indicate the
usefulness of such a functional.

\noindent As indicated in the literature, dynamic renormalization
group techniques, being useful and powerful, in many cases only
offers incomplete results, having no access to the strong coupling
phase [Barab\'asi \& Stanley, 1995; Wiese, 1998]. Hence, it is clear
the need of alternative ways to analyze the KPZ and related
problems, as for instance the self-consistent expansion [Katzav \&
Schwartz, 1999; Katzav, 2003]. The present results open new
possibilities of making non-perturbational studies for the KPZ
problem. For instance, through the analysis of long time mean values
of $h(\bar{x},t)$. In a similar way, it would be possible to obtain
correlations, and from them to extract information about scaling
exponents. Such study will be the subject of forthcoming work.

\bigskip

\noindent {\bf Acknowledgments} \smallskip

\noindent The author thanks R. Cuerno, H. Fogedby, J.M. L\'opez,
M.A. Mu\~{n}oz, L. Pesquera, J. Revelli, M.A. Rodriguez, R. Toral
for fruitful discussions and/or valuable comments. He also
acknowledges financial support from MEC, Spain, through Grant No.
CGL2007-64387/CLI; and the award of a {\it Marie Curie Chair}, from
the European Commission, during part of the development of this
work.

\bigskip

\noindent {\bf References} \smallskip

\noindent Ahlers V. \& Pikovsky A. [2002] ``Critical Properties of
the Synchronization Transition in Space-Time Chaos", \textit{Phys.
Rev. Lett.} \textbf{88}, 254101.

\noindent Ao P. [2004] ``Potential in stochastic differential
equations: novel construction", \textit{J. Phys. A} \textbf{37},
L25-L30.

\noindent Barab\'asi A.-L. \& Stanley H.E. [1995] \textit{Fractal
concepts in surface growth} (Cambridge U,P., Cambridge).

\noindent Bouzat S. \& Wio H.S. [1998] ``Nonequilibrium potential
and pattern formation in a three-component reaction-diffusion
system", \textit{Phys. Lett. A} \textbf{247}, 297-302.

\noindent Bouzat S. \& Wio H.S. [1999] ``Stochastic resonance in
extended bistable systems: The role of potential symmetry",
\textit{Phys. Rev. E} \textbf{59}, 5142-5149.

\noindent Castro M., et al. [2007] ``Generic equations for pattern
formation in evolving interfaces", \textit{New J. Phys.} \textbf{9},
102.

\noindent Colaiori F. \& Moore M.A. [2002] ``Numerical solution of
the mode-coupling equations for the Kardar-Parisi-Zhang equation in
one dimension", \textit{Phys. Rev. E} \textbf{65}, 017105.

\noindent Cross M.C. \& Hohenberg P.C. [1993] ``Pattern formation
outside of equilibrium", \textit{Rev. Mod. Phys.} \textbf{65},
851-1112.

\noindent de los Santos F., Romera E., Al Hammal O. \& Mu\~{n}oz
M.A. [2007] ``Critical wetting of a class of nonequilibrium
interfaces: A mean-field picture", \textit{Phys. Rev. E} \textbf{75},
031105.

\noindent Descalzi O. \& Graham R. [1994] ``Nonequilibrium potential
for the Ginzburg-Landau equation in the phase-turbulent regime",
\textit{Z. Phys. B} \textbf{93}, 509-513.

\noindent Edwards S.F. \& Schwartz M. [2002] ``Lagrangian
statistical mechanics applied to non-linear stochastic field
equations ", \textit{Physica A} \textbf{303}, 357-386.

\noindent Fogedby H.C. [2006] ``Kardar-Parisi-Zhang equation in the
weak noise limit: Pattern formation and upper critical dimension",
\textit{Phys. Rev. E} \textbf{73}, 031104.

\noindent Gammaitoni L., H\"anggi P., Jung P. \& Marchesoni F.
[1998] ``Stochastic resonance", \textit{Rev. Mod. Phys.}
\textbf{70}, 223-287.

\noindent Graham R. [1987], in {\it Instabilities and Nonequilibrium
Structures}, Eds.E.Tirapegui and D.Villaroel (D.Reidel, Dordrecht)

\noindent Graham R. \& Tel T. [1990] ``Steady-state ensemble for the
complex Ginzburg-Landau equation with weak noise", \textit{Phys.
Rev. A} \textbf{42}, 4661-4667.

\noindent Grinstein G., Mu\~noz M.A. \& Tu Y. [1996] ``Phase
Structure of Systems with Multiplicative Noise", \textit{Phys. Rev.
Lett.} \textbf{76}, 4376 - 4379.

\noindent H\"{a}nggi P. [1985], ``The functional derivative and its
use in the description of noisy dynamical systems" in
\textit{Stochastic Processes Applied to Physics}, Pesquera L. \&
Rodriguez M.A., Eds, (World Scientific, Singapore) pgs. 69-94.

\noindent Halpin-Healy T. \& Zhang Y.-Ch. [1995], \textit{Phys.
Rep.} \textbf{254}, 215-414.

\noindent Hayase, Y. [1997] ``Collision and Self-Replication of
Pulses in a Reaction Diffusion System", \textit{J. Phys. Soc. Jpn.}
\textbf{66}, 2584-2587.

\noindent Hentschel H.G.E. [1994] ``Shift invariance and surface
growth", \textit{J. Phys. A} \textbf{27}, 2269-2276.

\noindent Hohenberg P. \& Halperin B. [1977] ``Theory of dynamic
critical phenomena", \textit{Rev. Mod. Phys.} \textbf{49}, 435-479.

\noindent Iz\'us G., Deza R. \& Wio H.S. [1998] ``Exact
nonequilibrium potential for the FitzHugh-Nagumo model in the
excitable and bistable regimes", \textit{Phys. Rev. E} \textbf{58},
93-98.

\noindent Kardar M., Parisi G. \& Zhang Y.C. [1986] ``Dynamic
Scaling of Growing Interfaces", \textit{Phys. Rev. Lett.}
\textbf{56}, 889-892.

\noindent Katzav E. [2003] ``Self-consistent expansion results for
the nonlocal Kardar-Parisi-Zhang equation", \textit{Phys. Rev. E}
\textbf{68}, 046113.

\noindent Katzav E. \& Schwartz M. [1999] ``Self-consistent
expansion for the Kardar-Parisi-Zhang equation with correlated
noise", \textit{Phys. Rev. E} \textbf{60}, 5677-5680.

\noindent Katzav E. \& Schwartz M. [2004] ``Numerical evidence for
stretched exponential relaxations in the Kardar-Parisi-Zhang
equation", \textit{Phys. Rev. E} \textbf{69}, 052603.

\noindent Krug J. \& Halpin-Healy T. [1993] ``Directed polymers in
the presence of columnar disorder", \textit{J. Phys. I} \textbf{3},
2179-2198.

\noindent Kuramoto Y. [1978] ``Diffusion-Induced Chaos in Reaction
Systems", \textit{Suppl. Prog. Theor. Phys.} \textbf{64}, 346-367.

\noindent Lam C.-H. \& Shin F.G. [1998], ``Improved discretization
of the Kardar-Parisi-Zhang equation", \textit{Phys. Rev. E} \textbf{58},
5592.

\noindent Linz S.J., Raible M. \& H\"{a}nggi P. [2000], in
\textit{Stochastic Processes in Physics}, Freund J.A. \& P\"{o}schel
T., Eds., \textit{Lecture Notes in Physics}, \textbf{557}, 473-483
(Springer-Verlag, Berlin).

\noindent L\'opez J.M., Castro M. \& Gallego R. [2005] ``Scaling of
Local Slopes, Conservation Laws, and Anomalous Roughening in Surface
Growth", \textit{Phys. Rev. Lett.} \textbf{94}, 166103.

\noindent Ma K., Jiang J. \& Yang C.B. [2007] ``Scaling behavior of
roughness in the two-dimensional Kardar–Parisi–Zhang growth",
\textit{Physica A} \textbf{378}, 194-200.

\noindent Marsili M., Maritain A., Toigo F. \& Banavar J.R. [1996]
``Stochastic growth equations and reparametrization invariance",
\textit{Rev. Mod. Phys.} \textbf{68}, 963-983.

\noindent Medina E., Hwa T., Kardar M. \& Zhang Y.C. [1989]
``Burgers equation with correlated noise: Renormalization-group
analysis and applications to directed polymers and interface
growth", \textit{Phys. Rev. A} \textbf{39}, 3053-3075.

\noindent Mikhailov A.S. [1990], \emph{Foundations  of Synergetics
I} (Springer-Verlag, Berlin).

\noindent Montagne R., Hern\'andez-Garcia E. \& San Miguel M. [1996]
``Numerical study of a Lyapunov functional for the complex
Ginzburg-Landau equation", \textit{Physica D} \textbf{96}, 47-65.

\noindent Mukherji S. \& Bhattacharjee S.M. [1997] ``Nonlocality in
Kinetic Roughening", \textit{Phys. Rev. Lett.} \textbf{79},
2502-2505.

\noindent Nattermann T. \& Renz W. [1989] ``Diffusion in a random
catalytic environment, polymers in random media, and stochastically
growing interfaces", \textit{Phys. Rev. A} \textbf{40}, 4675-4681.

\noindent Newman T,J, \& Bray A.J. [1996], ``Strong coupling
behaviour in discrete Kardar–Parisi–Zhang equations", \textit{J.
Phys. A} \textbf{29}, 7917.

\noindent Pr\"{a}hofer M. \& Spohn H. [2004] ``Exact Scaling
Functions for One-Dimensional Stationary KPZ Growth", \textit{J.
Stat. Phys.} \textbf{115}, 255-279.

\noindent Ramasco J.J., L\'opez J.M. \& Rodriguez M.A. [2006]
``Interface depinning in the absence of an external driving force",
\textit{Phys. Rev. E} \textbf{64}, 066109.

\noindent Revelli J.A. \& Wio H.S. [2008], ``Some discretization
problems in the one dimensional KPZ equation", \textit{to be
submitted.}

\noindent Schwartz M. \& Edwards S.F. [2002] ``Stretched exponential
in non-linear stochastic field theories", \textit{Physica A}
\textbf{312}, 363-368.

\noindent Sivashinsky G. [1977] ``Nonlinear analysis of hydrodynamic
instability in laminar flames—I. Derivation of basic equations",
\textit{Acta Astronautica} \textbf{4}, 1177-1206.

\noindent Sun T., Guo H. \& Grant M. [1989] ``Dynamics of driven
interfaces with a conservation law", \textit{Phys. Rev. A}
\textbf{40}, 6763-6766.

\noindent Szendro I.G., L\'opez J.M. \& Rodriguez M.A. [2007]
``Localization in disordered media, anomalous roughening, and
coarsening dynamics of faceted surfaces", \textit{Phys. Rev. E}
\textbf{76}, 011603.

\noindent Tessone C. \& Wio H.S. [2007] ``Stochastic Resonance in an
Extended FitzHugh-Nagumo System: The role of Selective Coupling",
\textit{Physica A} \textbf{374}, 46-54.

\noindent Tong W.M. \& Williams R.S. [1994], ``Kinetics of surface
growth", \textit{Annu. Rev. Phys. Chem.} \textbf{45}, 401-438.

\noindent Tu Y., Grinstein G. \& Mu\~noz M.A. [1997] ``Systems with
Multiplicative Noise: Critical Behavior from KPZ Equation and
Numerics", \textit{Phys. Rev. Lett.} \textbf{78}, 274-277.

\noindent von Haeften B., Deza R. \& Wio H.S. [2000] ``Enhancement
of stochastic resonance in distribute systems due to selective
coupling", \textit{Phys. Rev. Lett.} \textbf{84}, 404-407.

\noindent von Haeften B., Iz\'us G., Mangioni S., S\'anchez A.D. \&
Wio H.S. [2004] ``Stochastic resonance between dissipative
structures in a bistable noise-sustained dynamics" , \textit{Phys.
Rev. E} \textbf{69}, 021107.

\noindent von Haeften B., Iz\'us G. \& Wio H.S. [2005] ``System Size
Stochastic Resonance between Dissipative Structures in a Bistable
Noise-induced Dynamics", \textit{Phys. Rev. E} \textbf{72}, 021101.

\noindent von Haeften B. \& Wio H.S. [2007] ``Optimal Range of a
Non-Local Kernel for Stochastic Resonance in Extended Systems",
\textit{Physica A} \textbf{376}, 199-207.

\noindent Wiese K.J. [1998] ``On the Perturbation Expansion of the
KPZ Equation", \textit{J. Stat. Phys.} \textbf{93}, 143-154.

\noindent Wio H.S. [1994], \textit{An Introduction to Stochastic
Processes and Nonequilibrium Statistical Physics} (World Scientific,
Singapore).

\noindent Wio H.S. [1997], ``Nonequilibrium Potential in
Reaction-Diffusion Systems", in \textit{4th Granada Lectures in
Computational Physics}, Eds Garrido P.L. \& Marro J.,
(Springer-Verlag), pg.135-193.

\noindent Wio H.S. [2007], ``Nonequilibrium Free Energy-Like
Functional for the KPZ Equation", arXiv:0709.4439, submitted to
\textit{Europhys. Lett.}.

\noindent Wio H.S., Bouzat S. \& von Haeften B. [2002] ``Stochastic
Resonance in  Spatially Extended Systems: the Role of Far From
Equilibrium Potentials", \textit{Physica A} \textbf{306C} 140-156.

\noindent Wio H.S. \& Deza R. [2007] ``Aspects of stochastic
resonance in reaction--diffusion systems: The
nonequilibrium-potential approach", \textit{Europ. Phys. J - Special
Topics} \textbf{146}, 111-126.

\end{document}